 \documentclass[preprint, review,3p]{elsarticle}

\usepackage{amssymb}

\usepackage[utf8]{inputenc}
\usepackage{amsmath}
\usepackage{soul}
\usepackage{color}
\biboptions{sort&compress}

\journal{Ultramicroscopy}

\begin{document}

\begin{frontmatter}

\title{Numerical modeling of specimen geometry for quantitative energy dispersive X-ray spectroscopy}

\author[label1]{W. Xu}
\address[label1]{Department of Materials Science and Engineering, North Carolina State University, Raleigh, NC 27695, USA.}

\author[label1]{J. H. Dycus}
\author[label1]{J. M. LeBeau\corref{cor1}}
\ead{jmlebeau@ncsu.edu}

\cortext[cor1]{Corresponding author}

\begin{abstract}

Transmission electron microscopy specimens typically exhibit local distortion at thin foil edges, which can influence the absorption of X-rays for quantitative energy dispersive X-ray spectroscopy (EDS). Here, we report a numerical, three-dimensional approach to model the geometry of general specimens and its influence on quantification when using single and multiple detector configurations. As a function of specimen tilt, we show that the model correctly predicts the asymmetric nature of X-ray counts and ratios. When using a single detector, we show that complex specimen geometries can introduce significant uncertainty in EDS quantification. Further, we show that this uncertainty can be largely negated by collection with multiple detectors placed symmetrically about the sample such as the FEI Super-X. Finally, based on guidance provided by the model, we propose methods to reduce quantification error introduced by the sample shape. The source code is available at https://github.com/subangstrom/superAngle.





\end{abstract}


\end{frontmatter}

\section{Introduction}

Energy dispersive X-ray spectroscopy (EDS) is widely used to determine the composition of materials in transmission electron microscopy \cite{WilliamCarter}.  Recent X-ray detector advances have dramatically enhanced collection efficiency through an enlargement of total detector area \cite{Zaluzec_2009_2} using either single or multiple detectors. In particular, these advances have been pioneered by four-quadrant \cite{SandiaTitan, Schlossmacher_2010superX} and dual, large-area detector configurations \cite{JEOL_dualEDS, FEI_dualEDS}. 
Advanced EDS systems also incorporate windowless silicon drift detectors (SDDs) \cite{Schlossmacher_windowlessSDD, Isakozawa2010}, which also increases the collection efficiency of low energy X-rays. These above developments have greatly improved the signal-to-noise ratio for two-dimensional mapping, offering many advantages, particularly elemental determination on an atom-column by atom-column basis \cite{Lu2014_1, Pinglu_2016, Chad_2015, Zanaga_2016, Houston_2016}.

\begin{figure}[h!]
\centering
\includegraphics[width=3.54in]{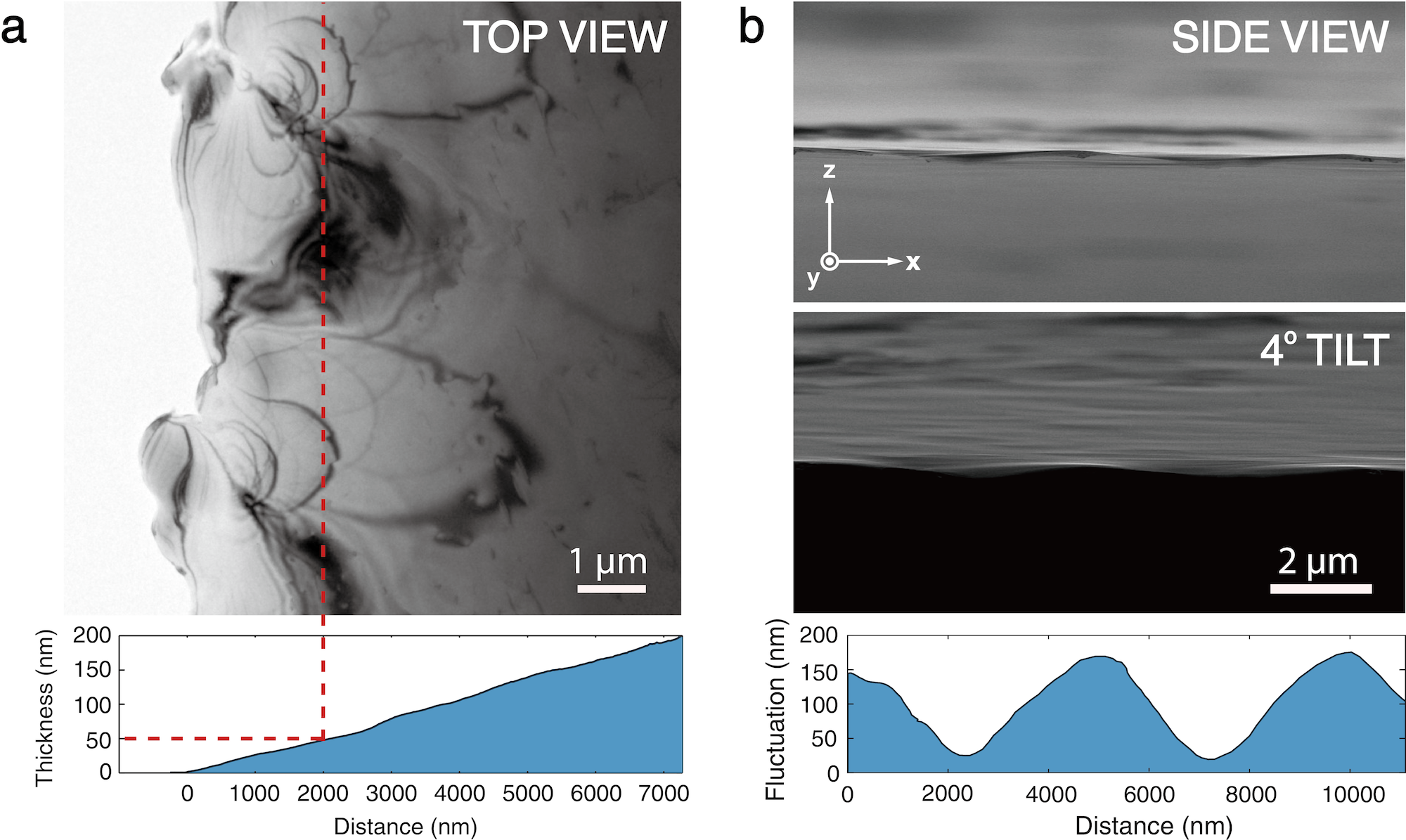}
\caption{(a) Bright-field TEM image of a typical Ni$_3$Al thin foil edge with corresponding thickness, calibrated from EELS. As highlighted by the (red) dashed lines, large local distortion is present in a typical thin region. (b, top) Secondary electron SEM images of the specimen and (b, bottom) sample distortion measured along the z direction.}
\label{fig:X1}
\end{figure}

Although the development of multi-detector collection systems has enabled new possibilities, the increased complexity has brought new challenges as well. One of which is X-ray shadowing by the specimen holder when using multiple detectors \cite{Yeoh2015, Slater_new_2016, Xu_2016, Kraxner_2017}. If only using a single detector, the holder shadowing can be avoided by tilting the sample towards that detector. Because this strategy is no longer an option when using multiple detectors, holder shadowing reduces the total detector solid angle for X-ray collection. Beyond holder shadowing effects, X-ray absorption calculations also become more complicated. As X-rays are collected from different orientations, local inhomogeneities of specimen shape become a critical factor for absorption correction. For example, local distortion is often seen in thin regions of TEM foils, impacting both the total X-ray counts collected as well as their ratios.  This is particularly problematic for material systems with strong absorption effects, such as Al-K absorption in Ni-Al alloys or materials containing light elements. A typical wedge-shaped Ni$_3$Al specimen, for example,  is presented in Figure \ref{fig:X1}a.  The bright-field TEM image shows bend contours across the thin edge, indicating local distortion. On average, these bend contours repeat every 5-7 $\mu$m along the thin edge, and extend to about 4-5 $\mu$m into the foil. These bend contours are representative of those found in general TEM thin foils, particularly after polishing and ion-milling.  

Distortion at the edge of thin foils can also be directly characterized via cross-sectional SEM imaging. Figure \ref{fig:X1}b shows the wave-like morphology at the thin edge with an oscillation amplitude of 70-80 nm. By tilting the sample 4$^{\circ}$ along the vertical axis, the locally distorted peaks and valleys can be viewed from a different vantage point.  In addition to bending along the  direction, further sample distortion is observed in the Y direction, perpendicular to horizontal, as shown in the SEM image with a 4$^{\circ}$ off-tilt. While surface indentions and foil folding are found at regions near the edge, the overall wedge thickness is linear as shown in Figure \ref{fig:X1}a. Beyond local distortion, the specimen surface is also inclined 5$^{\circ}$ relative to the support grid, resulting from slight misalignment of the sample when it was glued to the grid. 

For a typical sample, the regions of interest are often within the thin and distorted areas. Understanding how distorted specimen geometry affects X-ray absorption is thus critical for quantitative EDS analysis. As such, analytical solutions to correct for X-ray absorption have been proposed for flat plate and wedge-shaped samples when using single \cite{Goldstein_book, Zaluzec1984, WATANABE_1996} or multiple detectors \cite{Yang2014, Xu_2016}. In addition to absorption, holder shadowing and detector geometry have also been analytically included for electron tomography \cite{Burdet_2016,Slater_new_2016}. 

Towards realistic simulations of X-ray absorption, the authors have previously introduced a comprehensive numerical model to incorporate detector geometry, holder shadowing, and Be holder filtering. The combination of this model with dynamical electron scattering simulations enabled standard-less atomic resolution elemental quantification \cite{Chen_2016, Chen_2016_2}. This model, however, assumes that the specimen is a homogeneous, plate-shaped object to correct X-ray absorption. A universal model for complex specimen geometry is thus needed, particularly for arbitrarily shaped thin TEM foils. In particular, the impact of specimen geometry on quantification with multiple EDS detector is still unclear. Such understanding is fundamentally important for new approaches or strategies to reduce the quantification uncertainty due to specimen shape. 

In this paper, a three-dimensional, numerical approach is developed to model X-ray absorption correction for complex specimen geometries. The model predicts absorption correction factors using a mesh-based 3D object combined with the angular discretization of X-rays following the authors' prior work  characterizing multi-detector EDS \cite{Xu_2016}. With this approach, absolute X-ray counts and their ratios are predicted for realistic specimen shapes. Through a systematic study of distorted specimens, fundamental insights into the influence of complex sample geometry on quantitative EDS are provided.

\section{Materials \& Methods}

Due to the large attenuation coefficient of Al in Ni$_3$Al \cite{Goldstein_book}, Ni$_3$Al serves as an ideal model material to study the influence of distorted specimen geometry on quantitative EDS. A Ni$_3$Al super-alloy was thinned to electron transparency by wedge-polishing and low energy ion-milling. In addition, MgO particles were obtained by collecting the smoke of Mg burning in air. Formvar stabilized carbon grids, 75 mesh, were used to support the collected MgO particles. The X-ray parameters used to model these materials are provided in Table~\ref{tab:parameters}.

\begin{table}[tb]
	\caption{X-ray modeling parameters}
	\label{tab:parameters}
	\centering

	\begin{tabular}{l|ccc}
	\hline

	\hline
	 & \textbf{Cross-section (barn)} & \textbf{Fluorescence Yield} & \textbf{Detector Efficiency}  \cite{Schlossmacher_2010superX} \\
	\hline
		Ni & 255 \cite{Egerton_1994NiO} & 0.414 \cite{Egerton_1994NiO} & 0.980  \\
		Al & 2264  \cite{Zaluzec1984} & 0.0357 \cite{Bambynek_1972}  & 0.982 \ \\
		Mg & 2955  \cite{Egerton_EELS} & 0.032 \cite{Bambynek_1972}  & 0.981 \ \\
		O & 7980 \cite{Egerton_1994NiO} & 0.0085 \cite{Egerton_1994NiO}  & 0.923  \\

	\hline

	\hline
	\end{tabular}
\end{table}

An FEI Quanta 3D FEG instrument operated at 30 kV to determine the specimen geometry and bright-field TEM imaging was conducted using a JEM-2000FX at 200 kV. For STEM-EDS, a probe-corrected FEI Titan G2 TEM/STEM equipped with a four-quadrant FEI Super-X detector was used throughout the study and operated at 200 kV. The nominal EDS collection solid angle was 0.7 sr as reported elsewhere \cite{SandiaTitan}. Sample thickness was determined using the electron energy loss spectroscopy (EELS) log-ratio method implemented by the Digital Micrograph EELSTools suite \cite{Malis_EELS, Mitchell2005}.

For EDS experiments using Ni$_3$Al, spectra were acquired over an area of approximately 100 nm $\times$ 100 nm as a function of tilt about the X-axis ($\alpha$ tilt on FEI microscopes). For each sample tilt, an EDS spectrum was collected for a live time of 207 s for each individual detector. The probe current was measured to be ${7.83\times10^8}$ e/s via a CCD camera calibrated using the EELS drift tube method \cite{ishikawa_2014, Sang_2016}. The specimen position was -0.22 mm, 0.10 mm and 0.28 mm in x, y and z-coordinates relative to the holder, calibrated from a FIB sample grid. 

The specimen edge was parallel to the X-tilt axis of the holder with the thick region located closer to detectors 1 \& 2. The specimen thickness was kept at a constant value of 40 nm. The constant thickness approach reduced possible position variations, which potentially cause uncertainties in thickness determination for quantitative EDS. Strong electron channeling was avoided by selecting appropriate tilts about the X axis. To further reduce channeling, the sample was tilted -1$^\circ$ tilt about the Y direction.

To explore the influence of complex specimen geometry on materials containing light elements, X-rays from an MgO particle were collected with a live time of about 450 s (900 s total acquisition time) under a nominal 100 pA probe. The MgO particle had chamfered cubic morphology with dimensions, 205 nm $\times$ 195 nm $\times$ 190 nm, and was located near the middle area of a mesh grid. The corresponding specimen coordinates were -0.03 mm, 0.17 mm and 0.22 mm, respectively. The particle was tilted approximately 4$^\circ$ away from a zone axis to reduce the effect of channeling. 

\section{Description of the Model}


The numerical model accounts for X-ray absorption by first creating a three-dimensional representation of the sample via AutoCAD. The object is then imported into MATLAB and its volume is segmented into a mesh of tetrahedral units using the Partial Differential Equation Toolbox.  This process is the same as that used for mesh generation in finite element modeling \cite{FEM_mesh}. For example, to represent the specimen geometry shown in Figure \ref{fig:X1}a, the coordinates of a generated wedge-shaped object (wedge angle = 1.5$^\circ$) are displaced to match the as-measured wavy sample geometry. Then, an S-shaped sigmoid function is used to connect the wavy and non-distorted regions. The gradual inclination of the sample surface is also included. Note that other specimen features such as the local surface denting and folding are not considered below.


The specimen mesh is then used with the numerical model previously developed by the authors \cite{Xu_2016}.  The numerical model incorporates both the detector and holder geometries by angularly discretizing the collected X-rays. The effective detector solid-angle, $\Omega_{\mathrm{eff}}$ is determined by integrating over all discretized X-rays, $i$, collected from every sample slice along the depth direction, $t$, 

\begin{equation}
\Omega_{\mathrm{eff}}=\frac{1}{N}\sum_{t}\sum_{i}B_{i,\mathrm{frame}}B_{i,\mathrm{clip}}B_{i,\mathrm{grid}} \\
A_{i,\mathrm{spec}}(t)A_{i,\mathrm{carr}}d\theta d\varphi \sin\theta_i
\label{eq:model1}
\end{equation}

\noindent where $A_{i, \mathrm{spec}}$ and $A_{i, \mathrm{car}}$ represent the X-ray absorption factors from the specimen and Be specimen carrier, respectively.  In addition, the specimen holder frame $B_{i,\mathrm{frame}}$, sample securing clip $B_{i,\mathrm{clip}}$ and/or supporting grid $B_{i,\mathrm{grid}}$ may absorb nearly 100\%, i.e.~block, of the X-rays.

\begin{figure}[h!]
\centering
\includegraphics[width=3.00in]{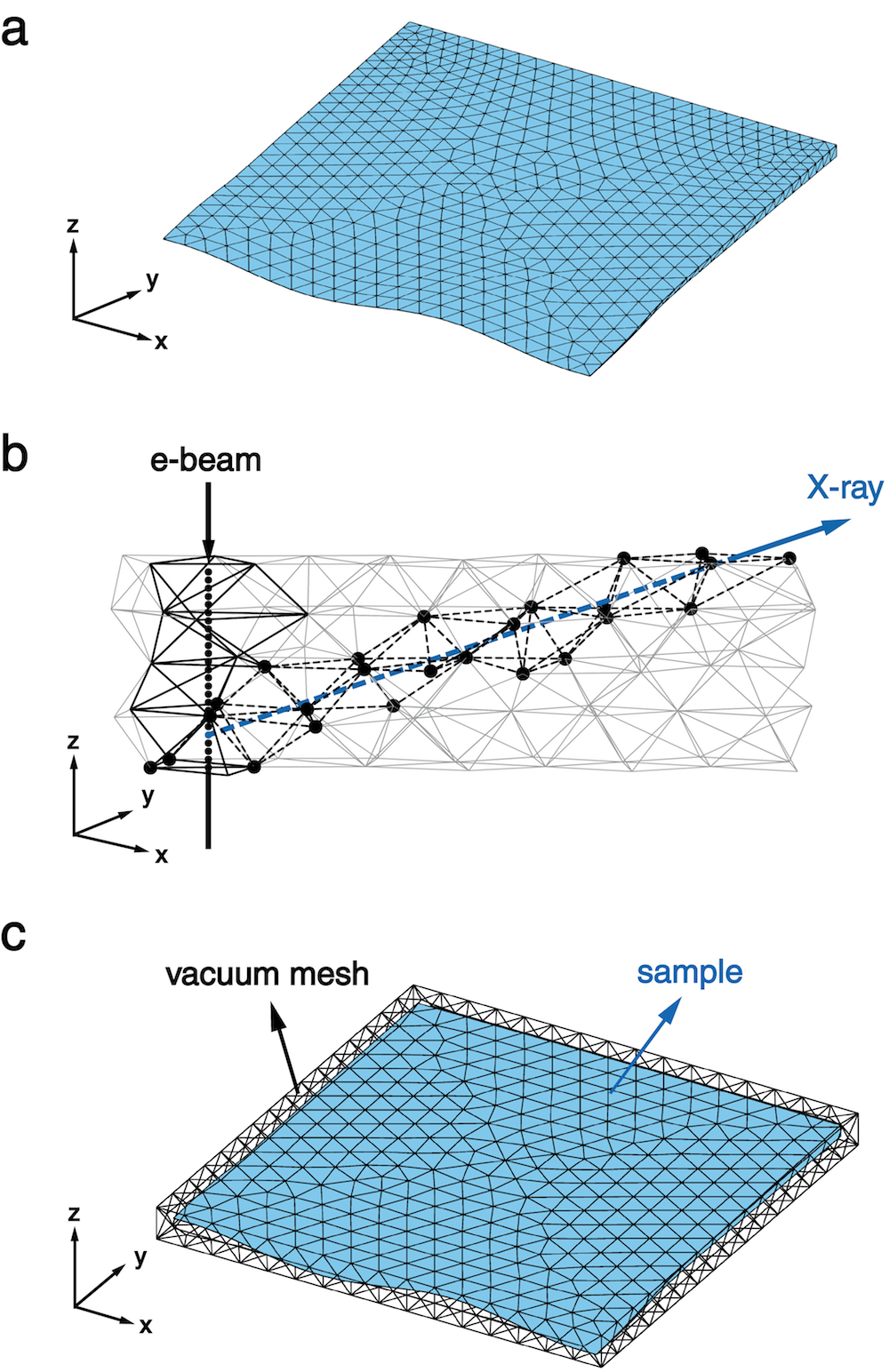}
\caption{(a) Three-dimensional model of the specimen shown in Figure \ref{fig:X1}. (b) The paths of the incident electron beam and an exiting X-ray  path inside the specimen are shown to intersect with the highlighted tetrahedra. (c) The vacuum is modeled to surround the sample.}
\label{fig:X2}
\end{figure}

The sample absorption term, 

\begin{equation}
A_{i,\mathrm{spec}}(t)=\exp(-l_{i,t}\mu\rho),
\label{eq:model2}
\end{equation}

\noindent depends on the distance, $l_{i,t}$, the X-ray travels inside the sample with specimen density $\rho$ and the mass attenuation coefficient $\mu$. $A_{i, \mathrm{spec}}$ is then calculated from the segmented specimen mesh, as detailed in the following steps. 

To calculate $l_{i,t}$ from the three-dimensional specimen mesh, the electron beam path through the sample model must be determined. First, the point where incoming electrons enter the sample top surface is found at the corresponding  tetrahedron using the M\"oller–Trumbore intersection algorithm \cite{Moller_1997}. The intersection of the beam path with the next tetrahedron is then calculated and repeated until the beam exits the specimen. Figure \ref{fig:X2}b provides an illustration of one such beam path and its intersections through the specimen (black tetrahedra). 

Once the electron path through the specimen is determined, it is further divided into $N$ thickness slices, as illustrated by the red dots in Figure \ref{fig:X2}b. The slice treatment makes it possible to calculate $l_{i,t}$ at different specimen depths, $t$, which is needed to incorporate the sample geometry in Equation \ref{eq:model1}. For each X-ray, $i$, at depth $t$, $l_{i,t}$ is determined by finding the distance between its origin (along the electron beam path) and its exit from the specimen.  The exit point is found using the same approach as tracking the beam path. The procedure is illustrated in Figure \ref{fig:X2}b, which shows the X-ray path $i$ (yellow line) and intersected tetrahedra (dashed tetrahedra). To accelerate the calculation, neighboring X-rays are evaluated to determine if they exit the same surface tetrahedron before initiating a new path. 

For a curved or more complicated sample geometry, X-rays exiting along particular trajectories can re-enter the specimen before reaching a detector, potentially multiple times. To take this scenario into account, a vacuum mesh surrounding the entire specimen is also created, as shown in Figure \ref{fig:X2}c. X-rays exiting the specimen surface are registered to the vacuum mesh and tracked until they completely leave the sample and vacuum mesh. The final distance $l_{i,t}$ is thus obtained by summing all X-ray paths that intersect the sample along X-ray path $i$, at thickness $t$.

\section{A simplified specimen model}

For a typical distorted specimen geometry (e.g.~Figure \ref{fig:X1}), the following three types of local sample distortion are observed,
\begin{enumerate}[(i)]
\item concave curvature
\item convex curvature
\item surface inclination
\end{enumerate}

\begin{figure*}[htbp]
\centering
\includegraphics[width=6.3in]{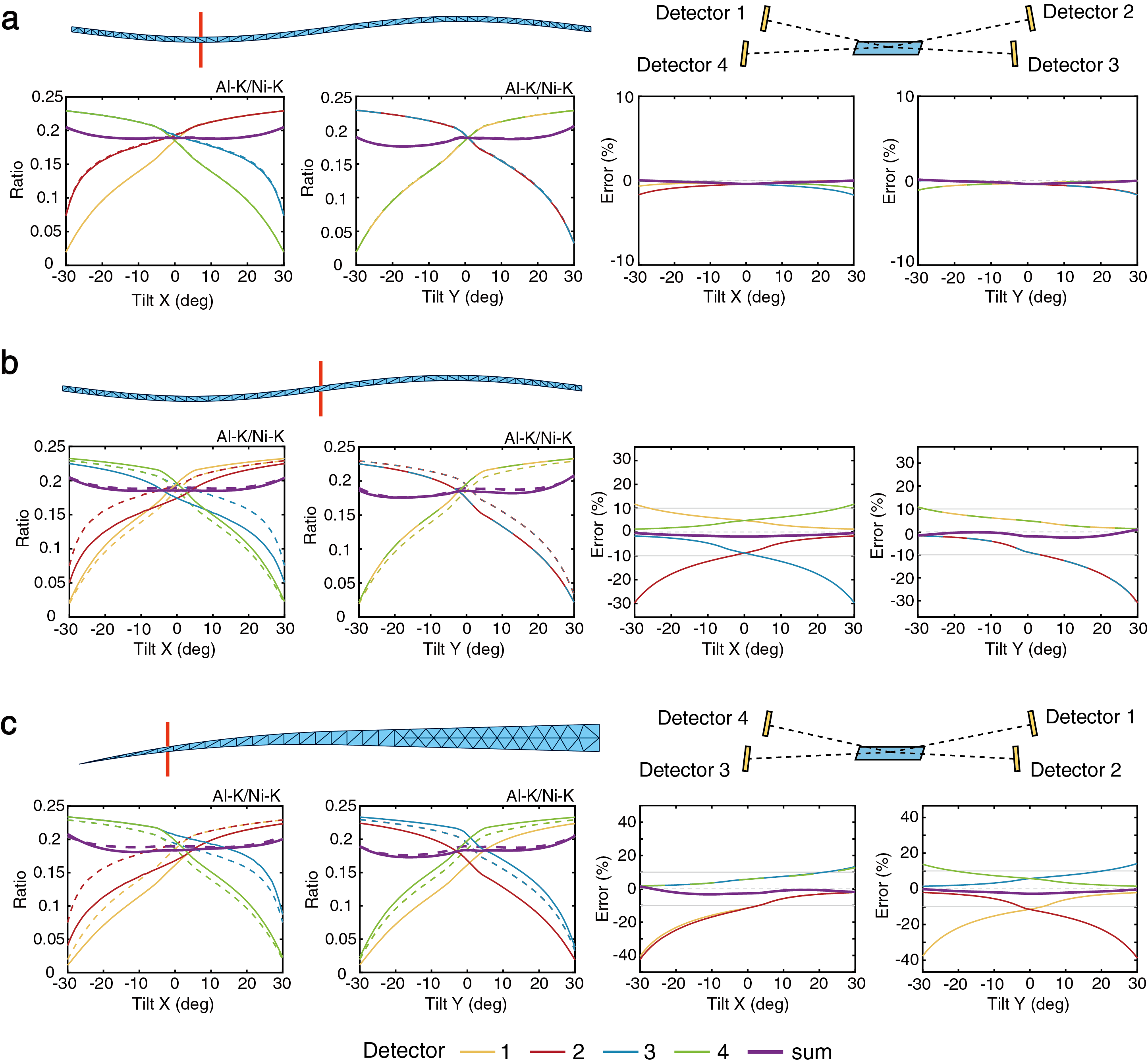}
\caption{Al-K/Ni-K ratios predicted assuming a sample region with (a) a concave dip, (b) surface inclined about the Y-axis, and (c) surface inclined about the X-axis. The X-rays measured at each detector and their sum are shown in each case for both the distorted sample (solid line) and a flat plate specimen (dash lines) as a function of X/Y-tilt. The quantification errors are shown at the right.}
\label{fig:X3}
\end{figure*}

For simplification, distortions can be modeled by considering a sine wave shaped  Ni$_3$Al sample.  The concave and surface inclination types of distortion are highlighted by the vertical markers in Figure \ref{fig:X3}a-b. The amplitude and spatial extent of the sine wave are chosen to correspond to the observations in Figure \ref{fig:X1}. Further, the simulations focus on a 50 nm thick region of the sample. To understand the influence of these geometries, absorption corrected Al-K/Ni-K intensity ratios for a flat, plate-shaped thin foil geometry (i.e.~the typically applied analytical solution) are used to calculate the quantification error relative to the distorted geometry (1 - C$_{Al,flat}$/C$_{Al, ideal\_distortion}$).  To reiterate, this quantification error is specifically referring to the uncertainty/inaccurate assumptions of specimen geometry, not other factors such as low counts or other unknowns.


Comparing the results from the distorted sample with undistorted flat thin film geometry, the center of the concave area (Figure \ref{fig:X3}a) shows a very small deviation of Al-K/Ni-K ratio. Minor compositional quantification errors are then introduced, which are limited to less than 2\% even with 30$^{\circ}$ of tilt along both tilt axes. Similarly, minor deviation is also found in the convex regions (not shown). 

The concave/convex distortions also cause surface inclination at their periphery. Such surface inclination, however, can result largely change the Al-K/Ni-K ratio from the flat shape geometry. The deviation becomes largest at local maxima or minima of the sine wave positions, where the specimen is inclined 4$^{\circ}$ about the Y-axis as shown in Figure \ref{fig:X3}b. With this deviation, quantification error will occur when simple flat specimen geometry is assumed. For each detector, the quantification error changes as a function of tilt. In particular, the quantification error for a single detector is greater than 5-10\% even at zero tilt, and it reaches 30\% at high tilts, inversely related to the tilt axis.

Consider the X-tilt case in Figure \ref{fig:X3}b -- detectors 1 and 4 show positive error whereas detectors 2 and 3 exhibit negative error. When X-rays collected from all detectors are added together, the net quantification error is mostly nullified in all cases. A similar behavior occurs when tilting along the Y-axis and the specimen surface is inclined about the X-axis. Furthermore, the magnitude of error is much larger for one detector pair than the other at high tilt angles. The quantification error counter-balancing is retained as a function of tilt because the contribution of each detector to the sum also changes with the tilt. The counter-balancing of quantification error is further explored in Section 6.  

\section{Direct comparisons to experiment}

\subsection{Ni$_3$Al thin foil} 
\label{sub:ni3al_thin_foil}

\begin{figure}[h!]
\centering
\includegraphics[width=3.54in]{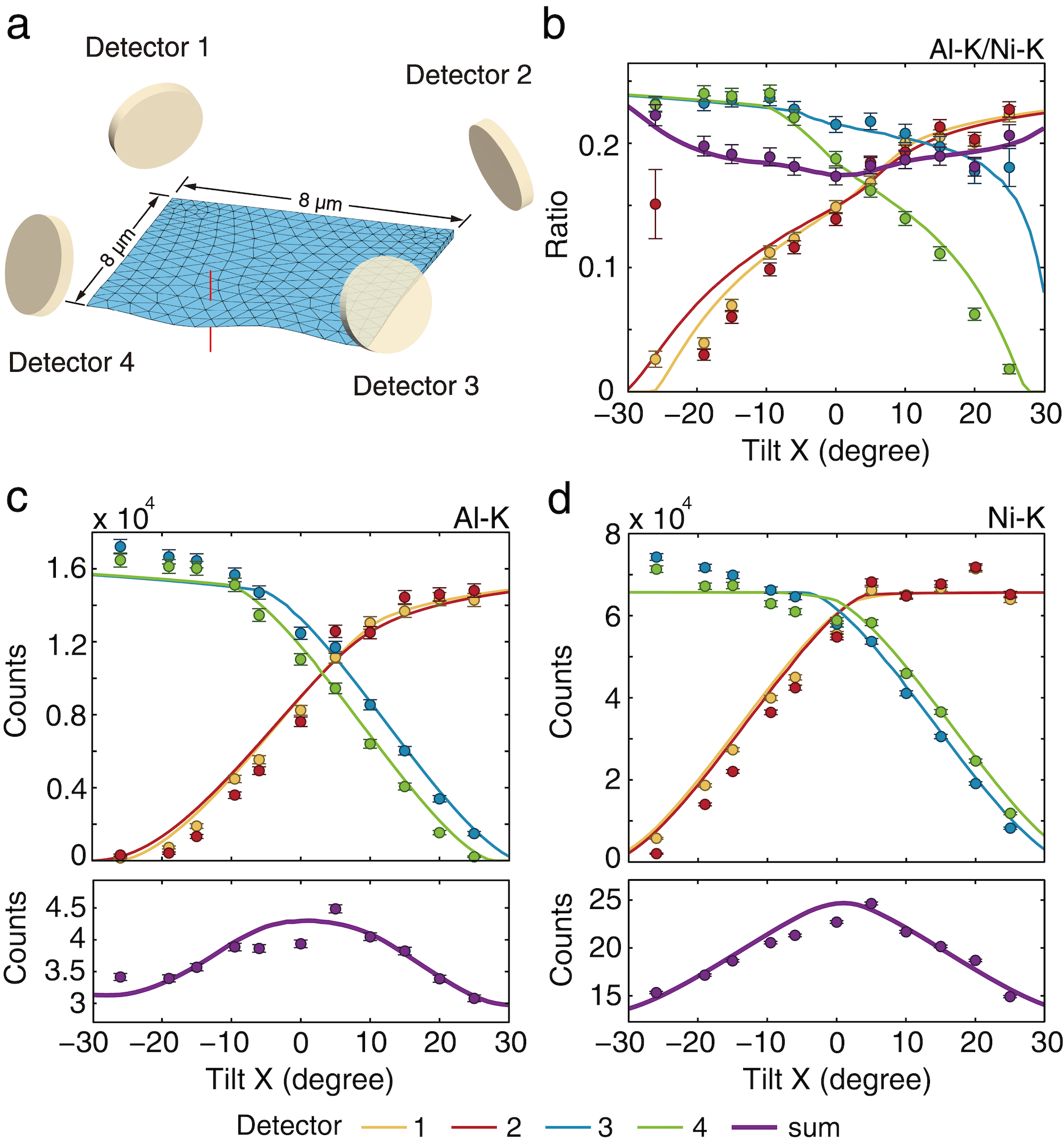}
\caption{(a) Three dimensional model of the specimen from experiment. For clarity, the illustrated mesh is less dense than the one used in simulations. Simulation location is indicated by the vertical line. Experiment (dots) and prediction (solid lines) of (b) Al-K/Ni-K ratio (c) Al-K and (d) Ni-K counts as a function of X-tilt.}
\label{fig:X4}
\end{figure}

Using the segmentation procedures discussed above, X-ray paths through the sample geometry of Ni$_3$Al from experiment (Figure \ref{fig:X1} and Figure \ref{fig:X4}a), are have been simulated. From these paths, the total Al-K and Ni-K X-ray counts, as well as their intensity ratio, are determined.  In each case, simulations (lines) agree with the experimental measurements (symbols) over an $\alpha$-tilt range of $\pm 30 ^{\circ}$ (Figures \ref{fig:X4}b-d). Two prominent trends are observed. First, the crossover of the Al-K/Ni-K ratios corresponding to detector pairs 1/2 and 3/4 systematically shift towards positive tilt angles. Second, the shift of the ratio for detectors 3 and 4 is very different compared with detectors 1 and 2. Further, the shift of the ratio is exaggerated at high specimen tilts as shown in Figure \ref{fig:X4}b. Also, note that from about -26$^{\circ}$ to -16$^{\circ}$ tilt in experiment, more X-ray counts are collected with detectors 3 and 4. This is likely attributed to residual electron channeling \cite{chen2015} 

The above deviations are mainly due to the change of specimen absorption resulting from the complex sample geometry.  For example, the Al-K counts shift toward positive tilt angles for detector combination 3 and 4, Figure \ref{fig:X4}c. Ni-K X-rays, on the other hand, are not significantly absorbed.  As a result, the Ni-K X-rays collected by all four detectors are more symmetric (Figure \ref{fig:X4}d). Only slight separation of Ni-K counts curves are observed due to variation in holder shadowing. Also, it is important to note the sum of all four detectors exhibits minimal changes as a function of tilt, which further highlights the advantage of combining signals from multiple detectors to reduce the influence of sample geometry. Overall, the agreement indicates that the proposed numerical model can correctly predict the X-rays collected for a distorted specimen. 

\begin{figure}[h!]
\centering
\includegraphics[width=3.00in]{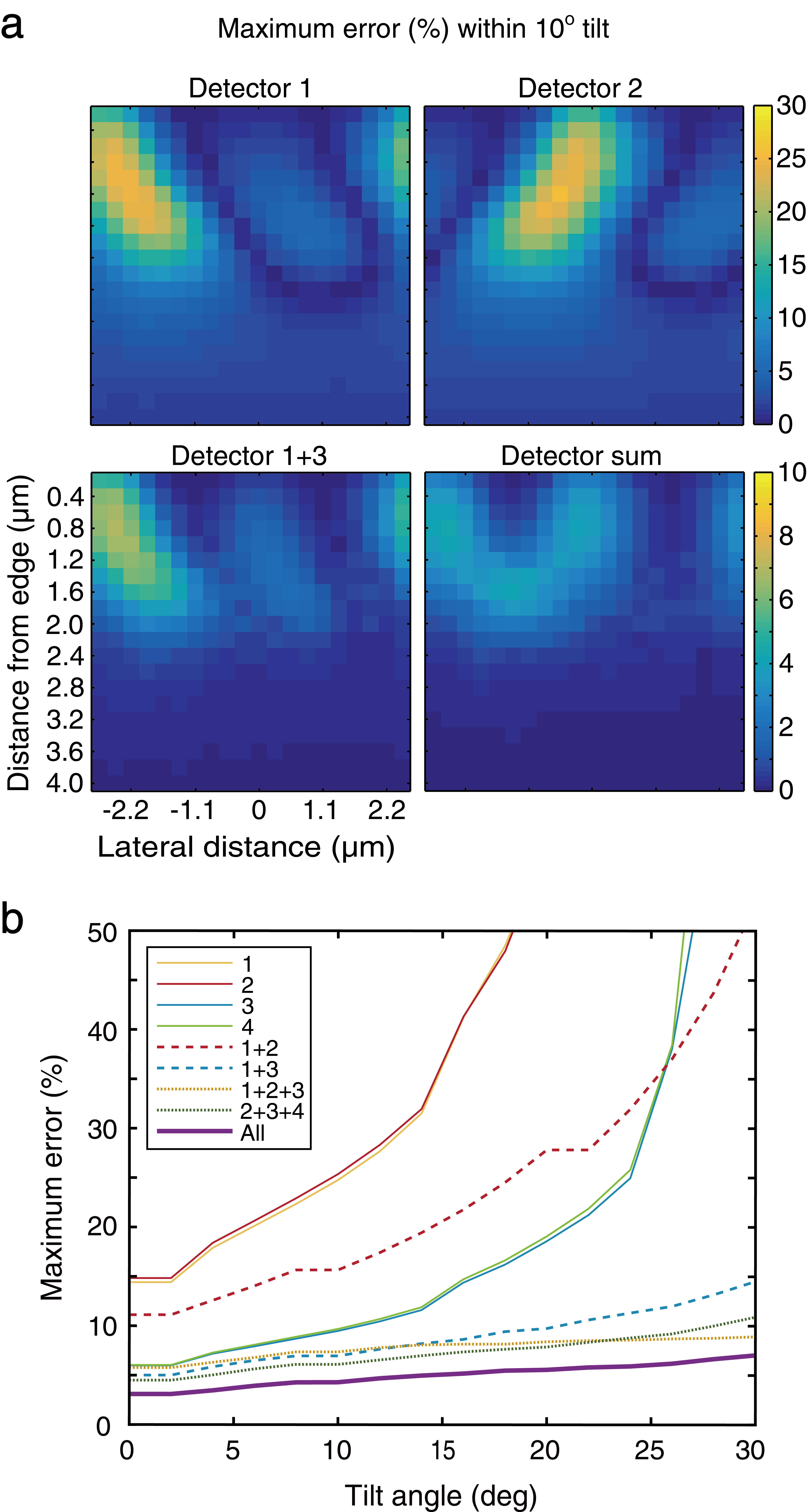}
\caption{Two-dimensional map of the maximum composition quantification error when using a simplified flat plate geometry for detector (a) 1 (b) 2 and (c)  and sum. (d) Plot of the maximum quantification error over the whole specific region as a function of tilt angle up to $\pm$30$^\circ$. Colors are used to indicate the individual detectors and their combination.}
\label{fig:X5}
\end{figure}

Based on the realistic model characterized from experiment (Figure \ref{fig:X4}a), the Al-K/Ni-K ratio is simulated across the specimen area, with thickness ranging from 5 to 105 nm. Comparing these results to simulation using a flat homogeneous specimen, two-dimensional maps of maximum quantification error within a tilt range of 10$^{\circ}$ for detector 1 are shown in Figure \ref{fig:X5}a. As expected from Figure \ref{fig:X3}, the error is highly dependent on local geometry with values as large as 25\% found at the upper left, which corresponds to the specimen edge with a large surface inclination. 

In contrast to detector 1, detector 2 shows the largest quantification error near the middle of the specimen. Although the specimen also has a large surface inclination on the right, the small error indicates that considering both specimen geometry and detector geometry is crucial. When signals from opposing detectors (180$^\circ$ azimuth angle) are added together, a much smaller error is observed. For example, adding the signals from detectors 1 and 3, the error is reduced to less than 7\% over the entire area. When all four detectors are added together, the maximum quantification error decreases to only 4\%.

While identifying the precise local distortion is not possible in many practical cases, it is worthwhile to estimate the maximum possible quantification error that could be introduced due to geometric uncertainties. To do so, the maximum quantification error is collected over each two-dimensional map as a function of specimen tilt as shown in Figure \ref{fig:X5}b, i.e.~the maximum in Figure \ref{fig:X5}a becomes a data point in Figure \ref{fig:X5}b. In general, the maximum quantification error increases with the tilt angle. 

For a single detector, quantification error up to 15\% is possible even when the sample is not tilted. Combining signal from multiple detectors, however, greatly suppresses the error as in Figure \ref{fig:X5}b. The addition of perpendicular detectors 1 and 2 (90$^\circ$ azimuth angle) reduces the error by at least 20\% compared to either individual detector. By adding two collinear detectors, such as 1 and 3, the deviation is dramatically reduced by nearly a factor of three. As more detectors are combined, the error is further reduced, with four detectors reducing quantification error to a minimum. Furthermore, the error becomes much less sensitive to specimen tilt, which is less than 7\% at a tilt angle of 30$^{\circ}$.


\subsection{MgO smoke nanocubes}

\begin{figure*}[h!]
\centering
\includegraphics[width=6.5in]{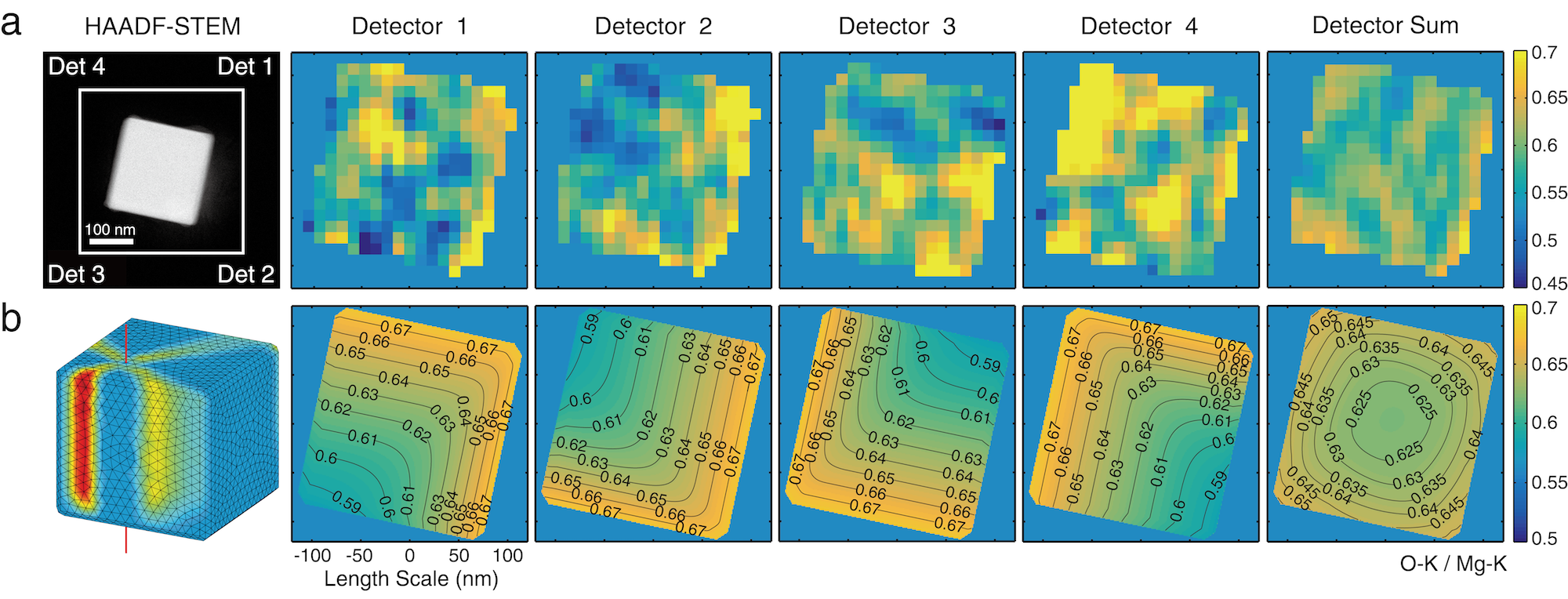}
\caption{O-K/Mg-K ratios from (a) experiment and (b) simulations using a MgO cube geometry. Relative positions between the sample and each detector are marked in the HAADF-STEM image.}
\label{fig:X6}
\end{figure*}

Beyond traditional foil samples, the model also enables elemental quantification of nanomaterials with complicated sample morphology. As an instructive example, MgO nanocubes with chamfered edges are considered. As a basis for comparison, Figure \ref{fig:X6}a shows O-K/Mg-K ratio maps over an entire MgO nanocube from experiment. Although the maps are fairly noisy, systematic variations of O-K/Mg-K ratio occur across the cube when using a single detector for collection. A larger ratio is observed near the specimen edge closest to each individual detector, while the ratio decreases when the specimen region is away from the detector. The O-K/Mg-K ratio is therefore site-dependent relative to each detector. 

The experiment trends are reproduced using the numerical model for the chamfered cube, as shown in Figure \ref{fig:X6}b. To understand the behavior,  shorter X-ray paths through the specimen occur for regions closer to the detector, resulting in a larger O-K/Mg-K ratio. Such variations can be significant -- 30-40\% for an individual detector in experiment, in part due to noise. A 15\% change of O-K/Mg-K ratio over whole specimen area is, however, also observed in the theoretical prediction. 

These results indicate that it becomes challenging to conduct accurate EDS quantification of nanomaterials using individual detectors, especially when light elements such as O (or C) are strongly absorbed by the specimen. In contrast, the ratio map is nearly flat when all four detectors are used. Less than 5\% variation is predicted across the cube from the numerical model as shown in Figure \ref{fig:X6}b. Near the center, the O-K/Mg-K variation is less than 2\% over a 100 $\times$ 100 nm region. This prediction is consistent with experiment, with an average O-K/Mg-K ratio predicted to be 0.63 and in excellent agreement with 0.60 from the experiment. As with the thin foil sample, these results indicate that it is a better strategy to use multiple detectors to balance the particle geometry induced quantification error. 



\section{Quantification Error Balancing}
\label{sec:counter-balance}

\begin{figure}[h!]
\centering
\includegraphics[width=3.45in]{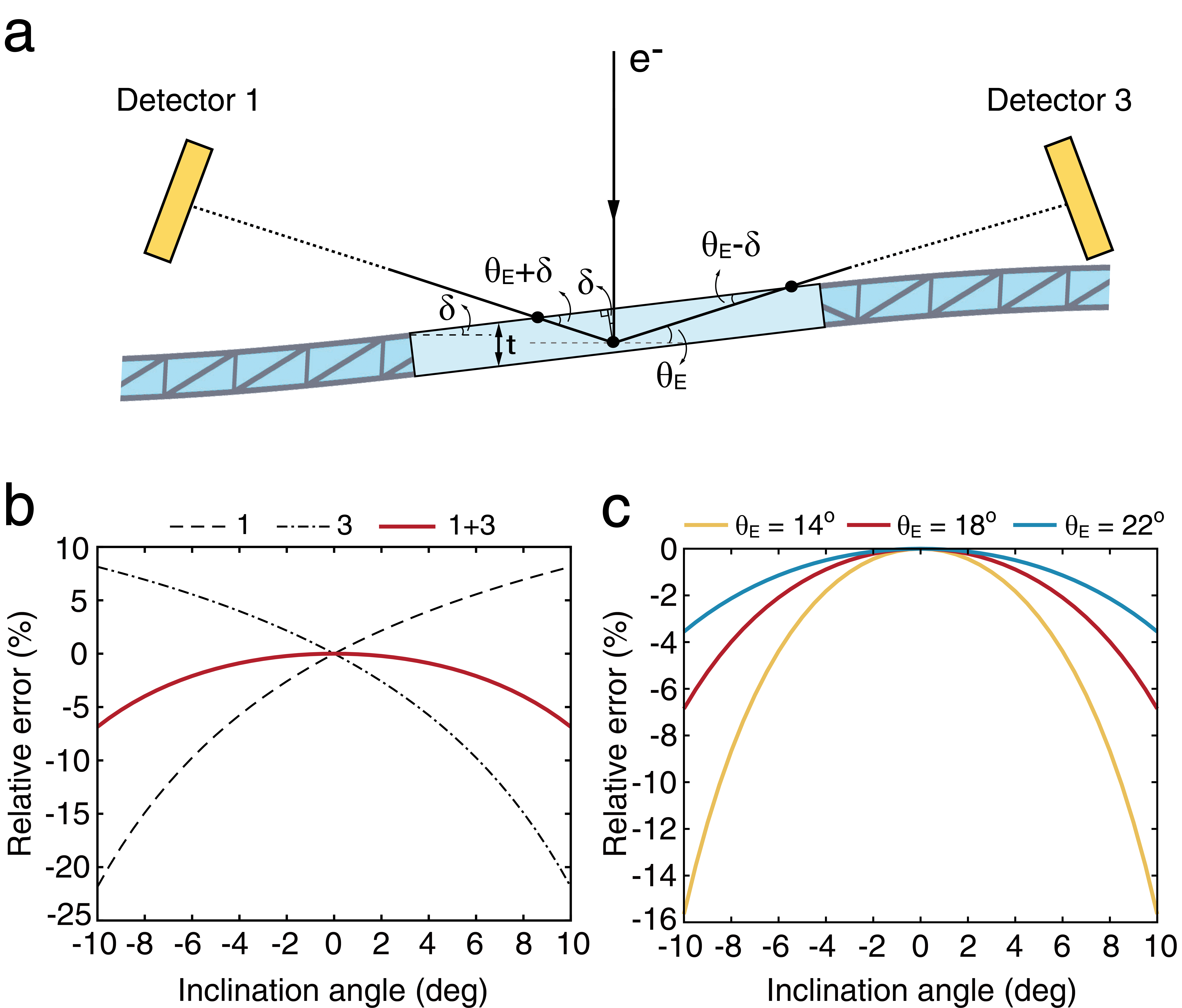}
\caption{(a) Path of an X-ray through a specimen inclined by $\delta$ towards detectors 1 and 3, which are in opposing positions (180$^\circ$ difference in the azimuth angle) located at an elevation angle of $\theta_E$. (b) Quantification error of the X-ray absorption for detectors 1/3 and their sum using a simplified analytical model compared to a non-inclined, plate shaped specimen. (c) Effect of the elevation angle on the estimated error.}
\label{fig:X7}
\end{figure}

To further explore quantification error balancing, consider an ideal model taking into account the following assumptions:
\begin{enumerate}[(i)]
\item  X-rays are generated from a specimen which is locally uniform but inclined by $\delta$ degrees along an axis perpendicular to the detectors
\item X-rays are collected by a pair of detectors at the same elevation $\theta_E$
\item the X-ray production rate is assumed constant through the thickness of the specimen
\item the two detectors (1 and 3) are collinear, i.e. 180$^\circ$ difference between their azimuth angles (Figure \ref{fig:X7}a);
\item the absorption correction factor (ACF) includes only two elements A and B, whose attenuation coefficients are $\left.\mu\right|_\mathrm{spec}^\mathrm A$ and $\left.\mu\right|_\mathrm{spec}^\mathrm B$
\item sample holder shadowing is neglected.
\end{enumerate}
In this model, only surface inclination is included as it is found to have the most dramatic influence on specimen absorption. 

For a specimen with thickness $t$ and material density $\rho$, the ACF for detector 1 in Figure \ref{fig:X7}a is  \cite{Goldstein_book}:
\begin{equation}
ACF_{1}=\bigg(\dfrac{\left.\mu\right|_\mathrm{spec}^\mathrm B}{\left.\mu\right|_\mathrm{spec}^\mathrm A}\bigg)\left(\dfrac{1-\exp\bigg[-\left.\mu\right|_\mathrm{spec}^\mathrm A\rho t\dfrac{\cos\delta}{\sin(\theta_E+\delta)}\bigg]}{1-\exp\bigg[-\left.\mu\right|_\mathrm{spec}^\mathrm B\rho t\dfrac{\cos\delta}{\sin(\theta_E+\delta)}\bigg]}\right)
\label{eq:an1}
\end{equation}
The exponential term in Equation \ref{eq:an1} is well approximated by a Taylor expansion to the third order,
\begin{equation}
\begin{split}
\exp\bigg[-\mu\rho t\dfrac{\cos\delta}{\sin(\theta_E+\delta)}\bigg]=& 1-\mu\rho t\dfrac{\cos\delta}{\sin(\theta_E+\delta)}+\dfrac{1}{2}\bigg(\mu\rho t\dfrac{\cos\delta}{\sin(\theta_E+\delta)}\bigg)^2\\
&-\dfrac{1}{6}\bigg(\mu\rho t\dfrac{\cos\delta}{\sin(\theta_E+\delta)}\bigg)^3
\end{split}
\label{eq:an2_1}
\end{equation}
Equation \ref{eq:an1} can be further approximated to the following,
\begin{equation}
ACF_\mathrm{1}=\dfrac{1-\dfrac{k_A}{2}\bigg(\dfrac{\cos\delta}{\sin(\theta_E+\delta)}\bigg)+\dfrac{k_A}{6}\bigg(\dfrac{\cos\delta}{\sin(\theta_E+\delta)}\bigg)^2}{1-\dfrac{k_B}{2}\bigg(\dfrac{\cos\delta}{\sin(\theta_E+\delta)}\bigg)+\dfrac{k_B}{6}\bigg(\dfrac{\cos\delta}{\sin(\theta_E+\delta)}\bigg)^2}
\label{eq:an2_2}
\end{equation}

\noindent where $k_A=\left.\mu\right|_\mathrm{Ni_3Al}^\mathrm{Al}\rho t$ and $k_B=\left.\mu\right|_\mathrm{Ni_3Al}^\mathrm{Ni}\rho t$. In many scenarios,  absorption for X-rays from one element is trivial. For example, consider a Ni$_3$Al 50 nm thick specimen, where $k_B$ is only 0.0023 while $k_A$ is 0.1497 for Al. Therefore by assuming $k_B$ is zero, Equation \ref{eq:an2_2} can be further simplified to,
\begin{equation}
ACF_\mathrm{1}=1-\dfrac{k_A}{2}\bigg(\dfrac{\cos\delta}{\sin(\theta_E+\delta)}\bigg)+\dfrac{k_A}{6}\bigg(\dfrac{\cos\delta}{\sin(\theta_E+\delta)}\bigg)^2
\label{eq:an3}
\end{equation}
Similarly, the absorption correction factor for the opposing detector 3 can be expressed as: 
\begin{equation}
ACF_\mathrm{3}=1-\dfrac{k_B}{2}\bigg(\dfrac{\cos\delta}{\sin(\theta_E-\delta)}\bigg)+\dfrac{k_B}{6}\bigg(\dfrac{\cos\delta}{\sin(\theta_E-\delta)}\bigg)^2
\label{eq:an4}
\end{equation}
When the sample surface is not inclined ($\delta=0$), the corresponding absorption correction factor is:
\begin{equation}
ACF_\mathrm{0}=1-\dfrac{k_B}{2}\bigg(\dfrac{1}{\sin\theta_E}\bigg)+\dfrac{k_B}{6}\bigg(\dfrac{1}{\sin\theta_E}\bigg)^2
\label{eq:an4_2}
\end{equation}

Based on Equations \ref{eq:an3}-\ref{eq:an4_2}, the error due to specimen geometry uncertainty is determined by calculating the deviation from a specimen without inclination, i.e. $\Delta$ACF$_\mathrm{1}$=ACF$_\mathrm{1}$-ACF$_\mathrm{0}$ or $\Delta$ACF$_\mathrm{3}$=ACF$_\mathrm{3}$-ACF$_\mathrm{0}$,
\begin{equation}
  \begin{cases}
         \Delta ACF_\mathrm{1}=\dfrac{k_A}{2\tan\theta_E}\bigg(1-\dfrac{k}{3\sin\theta_E}-\dfrac{k\cos\delta}{3\sin^2(\theta_E+\delta)}\bigg)\dfrac{\sin\delta}{\sin(\theta_E+\delta)}\\
         \\
         \Delta ACF_\mathrm{3}=-\dfrac{k_A}{2\tan\theta_E}\bigg(1-\dfrac{k}{3\sin\theta_E}-\dfrac{k\cos\delta}{3\sin^2(\theta_E-\delta)}\bigg)\dfrac{\sin\delta}{\sin(\theta_E-\delta)}\\
  \end{cases}
\label{eq:an5}
\end{equation}
The error from the sum of detectors 1 and 3 for this set of X-rays, i.e. ($\Delta$ACF$_\mathrm{1}$+$\Delta$ACF$_\mathrm{3}$)/2, is then:
\begin{equation}
\begin{split}
\Delta ACF_\mathrm{1+3}= & -\dfrac{k_A}{2\tan\theta_E}\bigg(1-\dfrac{k_A}{3\sin\theta_E}-\dfrac{2k_A\cos^2\delta\sin\theta_E}{\sin(\theta_E+\delta)\sin(\theta_E-\delta)}\bigg)\\
& \frac{\cos\theta_E\sin^2\delta}{\sin(\theta_E+\delta)\sin(\theta_E-\delta)}
\end{split}
\label{eq:an6}
\end{equation}

Since $\Delta$ACF$_\mathrm{1+3}$ results from the sum of a positive and negative term with similar pre-factor terms (Equation \ref{eq:an5}), the error is largely cancelled. As a result, $\Delta$ACF$_\mathrm{1+3}$ is mostly proportional to $\sin^2\delta$ for a small inclination angle $\delta$. This is much different than the single detector case whose $\Delta$ACF$_\mathrm{I}$ (or $\Delta$ACF$_\mathrm{3}$) is proportional to $\sin\delta$. Considering that Equations \ref{eq:an5} and \ref{eq:an6} share a comparable pre-factor term, and the $\sin^2\delta$ relationship in $\Delta$ACF$_\mathrm{1+3}$, the final error is much smaller than suggested from simply adding two arbitrary detectors. 

The effect of quantification error balancing is graphically shown in Figure \ref{fig:X7}b. The relative errors of (ACF$_\mathrm{1}$-ACF$_\mathrm{0}$)/ACF$_\mathrm{0}$, (ACF$_\mathrm{3}$-ACF$_\mathrm{0}$)/ACF$_\mathrm{0}$ and (ACF$_\mathrm{1-3}$-ACF$_\mathrm{0}$)/ACF$_\mathrm{0}$ are calculated assuming a 50 nm thick Ni$_3$Al specimen and a 18$^\circ$ detector elevation angle. For a typical inclination angle up to 5$^\circ$, less than 2\% error occurs by combining detectors. Even for an extreme inclination angle of 10$^\circ$, only about 7\% error value is expected. These results confirm the effectiveness of combining detectors to limit the influence of geometric distortion on quantification error. 

\begin{figure}[h!]
\centering
\includegraphics[width=3.30in]{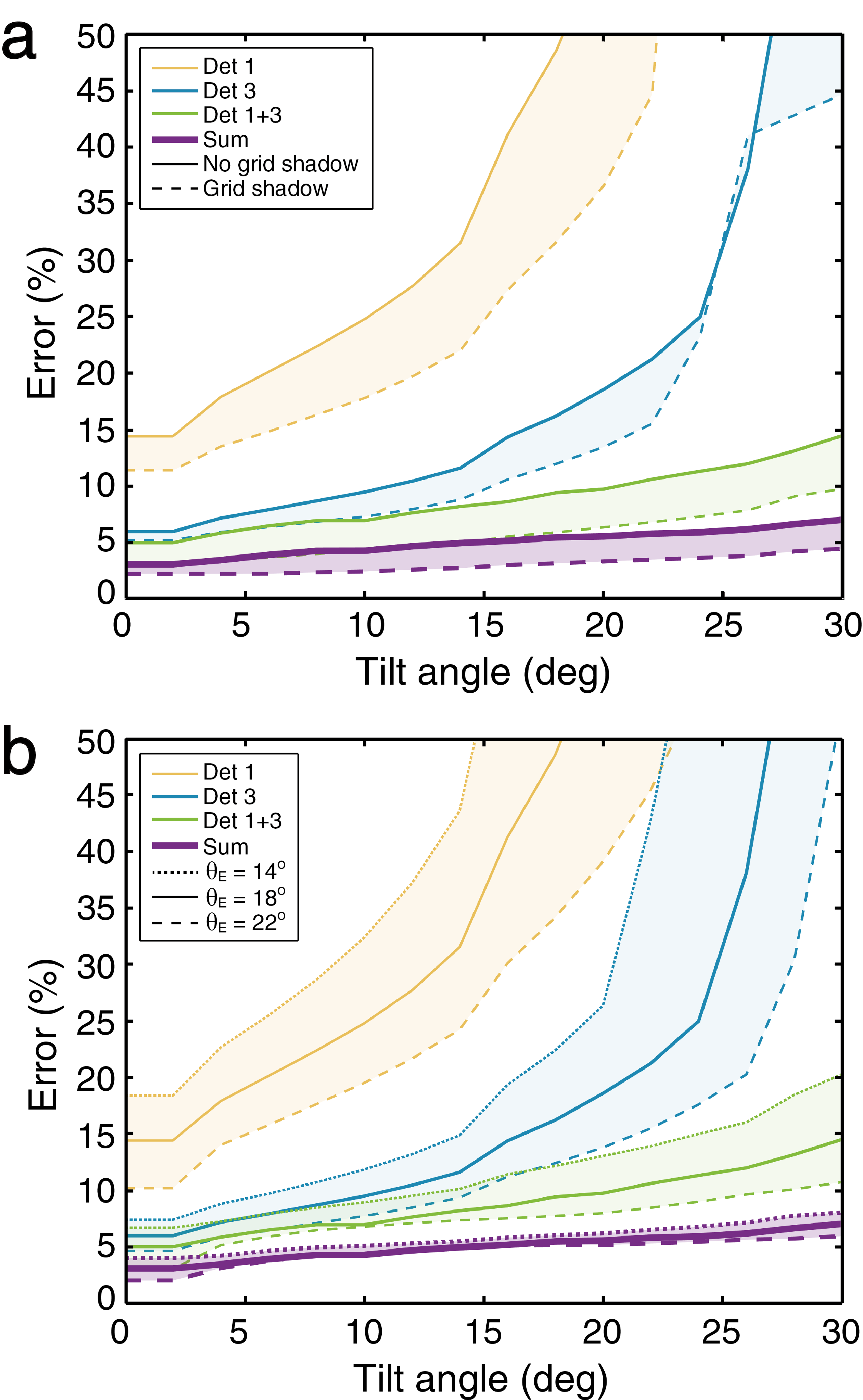}
\caption{(a) Maximum quantification error as a function of tilt angle ($\pm$30$^\circ$) for individual and multiple detectors with (solid lines) and without  grid shadowing (dashed lines). (b) Maximum quantification error for different detector elevation angles.}
\label{fig:X8}
\end{figure}

In addition to the collinear configuration, the analytical model can be readily developed for two perpendicular detectors (1 and 2).  Assuming that detector 1 has the same geometric relationship with the inclined specimen as in Figure \ref{fig:X7}a, the absorption correction term is then ACF$_\mathrm{1}$ in Equation \ref{eq:an5}. For detector 2, its ACF$_\mathrm{1}$ is identical to ACF$_\mathrm{0}$, i.e., it is not influenced by sample inclination. With these combined detectors, the error from adding detector I and II is then:

\begin{equation}
\Delta ACF_\mathrm{1+2}=\dfrac{1}{2}\dfrac{k_A}{2\tan\theta_E}\bigg(1-\dfrac{k}{3\sin\theta_E}-\dfrac{k\cos\delta}{3\sin^2(\theta_E+\delta)}\bigg)\dfrac{\sin\delta}{\sin(\theta_E+\delta)}\\
\label{eq:an8}
\end{equation}

While the combined error is reduced by half, the decrease is due to the non-geometric term of ACF$_\mathrm{2}$ and not from the error balancing discussed above. In other words, two perpendicularly positioned detectors are much less efficient at reducing quantification uncertainty than those that are opposite. For Super-X configuration, the quantification error balancing is always present since four detectors are placed 90$^\circ$ apart around the sample. Although the four-quadrant detector can be a good strategy to reduce the influence of sample geometry on quantification, an error of about 3-6\% is retained for realistically distorted geometry as seen in Figure \ref{fig:X5}e. 

\subsection{Quantification Error Reduction by blocking X-rays} 
\label{sub:grid}


The analytical solutions developed above also aid in understanding the results calculated from the numerical model. For example, consider the negative error observed in Figures \ref{fig:X3}c,d.  $\Delta$ACF$_\mathrm{1+3}$ (Equation \ref{eq:an6}) is always negative regardless of positive or negative $\delta$.  This indicates that specimen absorption will always be underestimated when assuming a non-inclined specimen geometry.  Such underestimation of specimen absorption can be further decreased via partially blocking the X-rays with smaller take-off angles, i.e.~ those with a larger specimen absorption component. 

Practically, blocking low take-off X-rays occurs by the sample holder or grid shadowing. While the above analytical model does not take X-ray shadowing into consideration, the effect from the specimen holder and grid shadowing can be demonstrated using the comprehensive numerical model. To do so in simulation, a 2 $\times$ 1 mm slot grid is considered on top of the specimen, and is compared with the error from neglecting the grid. In addition, the slot grid can also increase the relative height of the specimen to the holder, thus increasing shadowing by the specimen holder. As shown in Figure \ref{fig:X8}a, the maximum error across the entire specimen region, with same sample geometry as in Figure \ref{fig:X5}, is largely reduced. At large tilt angles, the maximum error of combining detectors 1+3 is reduced from 15\% to 10\% and  from 7\% to 5\% for the sum of all detectors.

\subsection{Effect of detector elevation angle}

According to Equation \ref{eq:an6}, other factors also directly contribute to quantification error balancing, such as $\mu\rho t_i$ and the elevation angle $\theta_E$. In particular, Equation \ref{eq:an6} predicts a significant influence of $\theta_E$ on the quantification error from the specimen geometry. The effect is visualized in Figure \ref{fig:X7}b, which compares the error between $\theta_E=14^\circ$, $18^\circ$, and $22^\circ$. Almost a factor of two difference in error occurs for a 4$^\circ$ change in $\theta_E$. The $\theta_E$ effect is further demonstrated in the more accurate numerical model shown in Figure \ref{fig:X8}b. The elevation angle uncertainty is particularly important for large area EDS detectors, which collect X-rays from a broad range of take-off angles\cite{Kraxner_2017}. This increases the quantification error for sample geometry uncertainty, and should be considered in the detector geometry design. Furthermore, the sample holder or grid shadowing can also partially alleviate the influence of $\theta_E$, since the X-rays from smaller $\theta_E$ are blocked. Such a strategy is, however, a trade-off as shadowing reduces the effective X-ray collection efficiency, which is the major benefit of having a large area detector.

\section{Conclusions}

A three-dimensional mesh-based model is established to calculate X-ray absorption coefficients for quantitative X-ray elemental analysis of specimens with complex geometry. In addition to the specimen, multiple detectors and holder shadowing/filtering are considered by the numerical model. Incorporating these contributions, predictions are shown to be in excellent agreement with experiment, both in terms of absolute X-ray counts and their ratios.

The model is also used to systematically investigate the quantification error resulting from complex sample geometry uncertainties when using single and multi-detector configurations. Based on the model results, it is found that local surface inclination has the greatest contribution to the quantification error, particularly when using a single X-ray detector. Further, the model explains how error can be readily counter--balanced when signals from at least two opposing detectors (180$^\circ$ apart) are combined. 

Finally, the model also provides opportunities to develop new strategies to reduce EDS quantification error due to sample geometry uncertainty.  For example, the model suggests that grid shadowing and an increased detector elevation angle can be used to reduce quantification error. Moving forward, it is anticipated that the flexibility of EDS numerical modeling can continue to aid in reducing uncertainty and can be used to develop improved sample holders optimized for detectors with large collection areas.

\section*{Acknowledgments}

The authors gratefully acknowledge the Air Force Office of Scientific Research (FA9550-14-1-0182) for support of this research. J.H.D. acknowledges support for this work by the National Science Foundation Graduate Research Fellowship (Grant DGE-1252376). This work was performed in part at the Analytical Instrumentation Facility (AIF) at North Carolina State University, which is supported by the State of North Carolina and the National Science Foundation (award number ECCS-1542015). The AIF is a member of the North Carolina Research Triangle Nanotechnology Network (RTNN), a site in the National Nanotechnology Coordinated Infrastructure (NNCI).

\section*{References}
\bibliographystyle{elsarticle-num}

\bibliography{refs}

\end{document}